\documentclass[twocolumn,eqsecnum,aps,showpacs]{revtex4}

\begin{document}   

\title{Superfield formulation of central charge anomalies\\
in two-dimensional supersymmetric theories with solitons}
 \author{K. Shizuya}
  \affiliation{Yukawa Institute for Theoretical Physics\\
Kyoto University,~Kyoto 606-8502,~Japan}

\begin{abstract} 

A superfield formulation is presented of the central charge anomaly in 
quantum corrections to solitons in two-dimensional theories with
$N=1$ supersymmetry. 
Extensive use is made of the superfield supercurrent, that places the
supercurrent $J^{\mu}_{\alpha}$, energy-momentum tensor $\Theta^{\mu\nu}$
and topological current $\zeta^{\mu}$ in a supermultiplet, to study
the structure of supersymmetry and related superconformal symmetry in the
presence of solitons.
It is shown that the supermultiplet structure of
$(J^{\mu}_{\alpha}, \Theta^{\mu\nu}, \zeta^{\mu})$ is kept exact 
while the topological current $\zeta^{\mu}$
acquires a quantum modification through the superconformal anomaly. 
In addition, the one-loop superfield effective action is explicitly
constructed to verify the BPS saturation of the soliton
spectrum as well as the effect of the anomaly.
\end{abstract}
 

\maketitle

\section{Introduction}
There has recently been considerable interest in quantum corrections to
solitons in supersymmetric theories in two dimensions.
As Witten and Olive~\cite{WO} pointed out long ago, in supersymmetric
theories with topological quantum numbers the supercharge algebra is
modified to include central charges and in certain theories the spectrum
of topologically stable excitations which saturate 
the Bogomol'nyi-Prasad-Sommerfield (BPS) bound classically is
determined exactly~\cite{DV}.
Solitons (or kinks) in two-dimensional theories with $N=1$
supersymmetry~\cite{VF}, in particular, are BPS saturated classically and
their masses are directly related to the central charges.  
Whether saturation continues at the quantum level in such theories,
however, has been a subject of unforeseen controversy. 
This is mainly because direct calculations of quantum corrections to the
soliton mass and/or central charge, as carried out by a number of
authors~\cite{KR,RN,NSNR,GJ,SVV,GLN,FN}, are a delicate task that requires
proper handling of regularization, renormalization and the boundary effect
simultaneously; see, e.g., refs.~\cite{SVV,FN}
for a survey on earlier research. 

It was shown eventually by Shifman, Vainshtein and Voloshin~\cite{SVV}
that the central charge is modified by a quantum anomaly so that, 
together with the quantum correction to the soliton mass, 
the BPS saturation is maintained at the quantum level.  A crucial element
in their analysis was the fact that they took advantage of
supersymmetry by starting with a superspace action with a supersymmetric
regularization prescription, although actual calculations were made with
component fields. It is therefore desirable to formulate the problem in a
manifestly supersymmetric framework.  Recently Fujikawa and van
Nieuwenhuizen~\cite{FN} have developed a superspace approach to the
central charge anomaly in two-dimensional theories. 
They have considered generalized local supersymmetry transformations on
the superfield and derived the central charge and related anomalies from
the associated Ward-Takahashi identities.

The purpose of this paper is to present a superfield formulation of the
central charge  and related anomalies in supersymmetric theories
from a somewhat different viewpoint.  
The key point in our approach is to first construct a superfield
supercurrent  that places the supercurrent $J^{\mu}_{\alpha}$, the
energy-momentum tensor $\Theta^{\mu\nu}$ and the topological current
$\zeta^{\mu}$ in a supermultiplet.
The  superspace Noether theorem for the supercurrent and related
superconformal currents is derived from the conservation law
of the superfield supercurrent, and possible quantum anomalies
are identified to arise from potentially anomalous products that take the
form of (fields)$\times$(equations of motion) which, when properly
regularized, could yield nonzero results.
We take full advantage of supersymmetry by use of the superspace
background-field method. 
It will be shown that the supercharge algebra, or equivalently the
supermultiplet structure of
$(J^{\mu}_{\alpha}, \Theta^{\mu\nu}, \zeta^{\mu})$, is kept exact 
while the topological current $\zeta^{\mu}$ acquires a quantum
modification at the one-loop level through the superconformal anomaly. 
In addition, the one-loop superfield effective action is explicitly
constructed to verify the anomaly and BPS saturation of the soliton
spectrum.

In Sec.~II we review some basic features of supersymmetric
theories with topological charges in two dimensions. In Sec.~III we
construct a superfield supercurrent and related superconformal
currents, and examine their conservation laws. 
In Sec.~IV we study the quantum anomalies by use of the
superspace background-field method. 
In Sec.~V we calculate the one-loop effective
action and examine its consequences. 
 Section~VI is devoted to a summary and discussion.

\section{N=I supersymmetry in two dimensions}

The $N=1$ superspace consists of points labeled by the space-time
coordinates
$x^{\mu}=(x^{0},x^{1})$ and two real Grassmann coordinates
$\theta_{\alpha}=(\theta_{1},\theta_{2})$; we denote
$z=(x^{\mu},\theta_{\alpha})$ collectively. Throughout this paper we adopt
the superspace notation of Ref.~\cite{SVV}.  For the Dirac matrices we
use the Majorana representation
\begin{equation}
\gamma^{0}=\sigma_{2}, \gamma^{1}=i\sigma_{3}
\end{equation}
with the charge-conjugation matrix $C_{\alpha \beta}=
-(\gamma^{0})_{\alpha \beta}=i\epsilon_{\alpha \beta}$ and 
$\epsilon_{12}=1$ so that for a charge-conjugate spinor
$\psi^{C}_{\alpha}\equiv C_{\alpha
\beta}\bar{\psi}_{\beta}=\psi_{\alpha}^{\dag}$ with
$\bar{\psi}=\psi^{\dag}\gamma^{0}$; 
accordingly Majorana spinors such as 
$\theta_{\alpha}$ are real spinors.

The real superfield $\Phi (z) \equiv \Phi (x,\theta)$ consists of a scalar
field $\phi (x)$ and a real spinor field $\psi_{\alpha} (x)$, along with
a real auxiliary field $F(x)$, with the expansion
\begin{equation}
\Phi (x,\theta) = \phi (x) + \bar{\theta}\psi (x) 
+ {1\over{2}}\,\bar{\theta}\theta\, F(x),
\end{equation}
where $\bar{\theta}\equiv \theta\gamma^{0} 
= i(\theta_{2}, -\theta_{1})$, 
$\bar{\theta}\theta=\bar{\theta}_{\alpha}\theta_{\alpha}=
-2i\theta_{1}\theta_{2}$ and $\bar{\theta}\psi\equiv
\bar{\theta}_{\alpha}\psi_{\alpha} = \bar{\psi}_{\alpha}\theta_{\alpha}$.
Translations in superspace
\begin{equation}
x^{\mu} \rightarrow x^{\mu} + i\bar{\xi} \gamma^{\mu} \theta ,\ \ 
\theta_{\alpha} \rightarrow \theta_{\alpha} +\xi_{\alpha}
\end{equation}
generate the variations of the superfield,
\begin{equation}
\delta \Phi (x,\theta) 
= \bar{\xi}_{\alpha} {\cal Q}_{\alpha}  \Phi (x,\theta)
\end{equation}
with 
${\cal Q}_{\alpha}= \partial/\partial \bar{\theta}_{\alpha}
+ i(\gamma^{\mu}\theta)_{\alpha}\partial_{\mu}$.
The component fields thereby undergo the supersymmetry transformations
\begin{eqnarray}
\delta \phi (x) &=& \bar{\xi}\psi (x), 
\delta \psi_{\alpha} (x) = -i (\gamma^{\mu}\xi)_{\alpha}
\partial_{\mu}\phi(x) + \xi_{\alpha} F(x), \nonumber\\
\delta F(x) &=& -i \bar{\xi}\gamma^{\mu} \partial_{\mu}\psi(x).
\label{susycomponent}
\end{eqnarray}

We consider a supersymmetric model described by a superfield
action of the form~\cite{VF}
\begin{equation}
 S = \int d^{2}x\, d^{2}\theta\, \left\{ {1\over{4}}\,
(\bar{D}_{\alpha}\Phi)\, D_{\alpha}\Phi + W(\Phi)
\right\}
\label{sfaction}
\end{equation}
with $\int d^{2}\theta\, {1\over{2}}\,\bar{\theta}\theta =1$;
the superpotential $W(\Phi)$ is a polynomial in $\Phi(x,\theta)$.
Here the spinor derivatives 
\begin{equation}
D_{\alpha}\equiv {\partial\over{\partial \bar{\theta}_{\alpha}}}
- (\slash\!\!\! p\, \theta)_{\alpha}
\end{equation}
with $\slash\!\!\! p \equiv \gamma^{\mu}p_{\mu}$ and
$p_{\mu}=i\partial_{\mu}$, or their conjugates
\begin{equation}
\bar{D}_{\alpha}\equiv D_{\beta}(\gamma^{0})_{\beta \alpha}
= -\partial/\partial \theta_{\alpha}  
+ (\bar{\theta} \slash\!\!\! p)_{\alpha} ,
\end{equation}
anticommute with supertranslations, $\{D_{\alpha}, {\cal Q}_{\beta}\}=0$,
and obey 
\begin{equation}
\{D_{\alpha}, D_{\beta}\} 
= 2\, (\slash\!\!\! p \gamma^{0})_{\alpha\beta};
\end{equation}
see Appendix A for some formulas related to $D_{\alpha}$.
In component fields the action reads 
$S = \int d^{2}x\, {\cal L}$ with
\begin{equation}
 {\cal L} =  {1\over{2}}\, \{\bar{\psi} i \slash\!\!\!\partial \psi 
+ (\partial_{\mu}\phi)^{2} + F^{2}\}  + FW'(\phi) 
-  {1\over{2}}\, W''(\phi)\, \bar{\psi}\psi ,
\label{modelcomponent}
\end{equation}
where $W'(\phi)=dW(\phi)/d\phi$, etc.  Eliminating the auxiliary field $F$
from ${\cal L}$, i.e., setting $F \rightarrow -W'(\phi)$, yields the
potential term $ -{1\over{2}}\, [W'(\phi)]^{2}$.

The superpotential $W(\Phi)$ may be left quite general in form.  
For definiteness we take $W(\Phi)$ to be odd in $\Phi$ and to have more
than one extrema  with $W'(\Phi)=0$ so that the model supports 
topologically stable solitons interpolating between different
degenerate vacua at spatial infinities
$x^{1} \rightarrow \pm \infty$. 
A simple choice~\cite{VF}
\begin{equation}
W(\Phi)= {m^{2}\over{4\lambda}}\, \Phi 
- {\lambda\over{3}}\, \Phi^{3}
\label{WZmodel}
\end{equation}
corresponds to the two-dimensional version of the Wess-Zumino
model~\cite{WZ}, and supports a classical static kink solution 
\begin{equation}
\phi^{\rm kink} (x) = (m/2\lambda)\tanh(mx^{1}/2)
\label{WBkink}
\end{equation}
interpolating between the two vacua with 
$\langle \phi \rangle_{\rm vac}=\pm m/(2\lambda)$ at $x^{1}=\pm \infty$.
Similarly, the supersymmetric sine-Gordon model with 
$W(\Phi)= m v^{2} \sin (\Phi/v)$ also supports solitons.

Associated with the supertranslations~(\ref{susycomponent}) of the
component fields is the Noether supercurrent
\begin{equation}
J^{\mu}_{\alpha} =
(\partial_{\nu}\phi)(\gamma^{\nu}\gamma^{\mu} \psi)_{\alpha}
-iF\, (\gamma^{\mu} \psi)_{\alpha}.
\label{Jmualpha}
\end{equation}
The conserved supercharges
\begin{equation}
Q_{\alpha} = \int dx^{1} J^{0}_{\alpha}
\end{equation}
generate, within the canonical formalism (and by naive use of equations
of motion), the transformation law of the supercurrent~\cite{SVV} 
\begin{equation}
i[\bar{\xi}_{\beta} Q_{\beta}, J^{\mu}_{\alpha}]
= 2\,(\gamma_{5}\xi)_{\alpha} \zeta^{\mu}
-2i(\gamma_{\lambda}\xi)_{\alpha} \Theta^{\mu
\lambda},
\label{deltaJ}
\end{equation}
where 
\begin{equation}
\Theta^{\mu \lambda} = {i\over{2}}\, \bar{\psi} 
\gamma^{\mu}\partial^{\lambda }\psi+
\partial^{\mu}\phi\partial^{\lambda}\phi 
- {1\over{2}}\, g^{\mu \lambda}\{(\partial_{\nu}\phi)^{2} 
- F^{2}\}   
\end{equation}
is the energy-momentum tensor and 
\begin{equation}
\zeta^{\mu} = -\epsilon^{\mu\nu} F \partial_{\nu}\phi 
= \epsilon^{\mu\nu} \partial_{\nu}W(\phi)
\label{zetanaive}
\end{equation}
is the topological current.
Here $\gamma_{5}\equiv \gamma^{0}\gamma^{1} = -\sigma_{1}$;
note the matrix identities specific to 1+1 dimensions,
\begin{equation}
\gamma^{\mu}\gamma_{5}=-\epsilon^{\mu \nu}\gamma_{\nu},\
\gamma^{\mu}\gamma^{\nu}= g^{\mu \nu} + \epsilon^{\mu \nu}\gamma_{5},
\end{equation} 
with $\epsilon^{01}=1$.

From Eq.~(\ref{deltaJ}) follows the supersymmetry algebra~\cite{WO,SJG}
\begin{equation}
\{Q_{\alpha}, \bar{Q}_{\beta}\} = 
2(\gamma_{\lambda})_{\alpha\beta}\, P^{\lambda}
+ 2i(\gamma_{5})_{\alpha\beta}\, Z
\label{superchargealgebra}
\end{equation}
where $P^{\lambda}=\int dx^{1} \Theta^{0\lambda}$ is the total
energy and momentum and
$Z$ is the central charge
\begin{equation}
Z = \int dx^{1} \zeta^{0}
= W(\phi)\vert_{x^{1}=\infty}- W(\phi)\vert_{x^{1}=-\infty}.
\label{Znaive}
\end{equation}
The $N=1$ superalgebra thus gets centrally extended in the presence of
solitons~\cite{WO}.

\section{Superfield supercurrent}

The transformation law~(\ref{deltaJ}) implies that the currents 
$J^{\mu}_{\alpha}$, $\zeta^{\mu}$ and $\Theta^{\mu \lambda}$ form a
supermultiplet. 
It is possible to construct a superfield supercurrent that exposes this
multiplet structure.
To this end note first that  $J^{\mu}_{\alpha}$ has dimension
$3/2$ in units of mass while 
$\Theta^{\mu \lambda}$ and $\zeta^{\mu}$ have dimension 2.
The supercurrent should thus be  a spinor-vector superfield (of dimension
3/2) that contains  $J^{\mu}_{\alpha}$ as
the lowest component, in sharp contrast to those in four
dimensions~\cite{FZ} that are vector superfields $\sim V_{\alpha
\dot{\alpha }}$.
It is quadratic in $\Phi$ and contains three $D$'s. The result is 
\begin{equation}
{\cal V}^{\mu}_{\alpha} =
-i(D_{\alpha}\bar{D}_{\lambda}\Phi)\, 
(\gamma^{\mu})_{\lambda\beta}\,D_{\beta}\Phi.
\label{sfcurrent}
\end{equation}
For formal consideration it is more appropriate to define 
${\cal V}^{\mu}_{\alpha}$ by the equivalent symmetrized form
\begin{equation}
-{i\over{2}}\left\{D_{\alpha}\bar{D}_{\lambda}\Phi \cdot 
(\gamma^{\mu})_{\lambda\beta}\,D_{\beta}\Phi 
+ \bar{D}_{\lambda}\Phi \cdot
(\gamma^{\mu})_{\lambda\beta}\,D_{\alpha}D_{\beta}\Phi \right\},
\end{equation}
but we shall use the notation~(\ref{sfcurrent}) with such
symmetrization understood from now on.  

The current ${\cal V}^{\mu}_{\alpha}$ is a real spinor superfield;
note, in this connection, that $(D_{\alpha})^{\dag} = -D_{\alpha}$ 
so that $D_{\alpha}\Phi$, e.g., is a real spinor superfield.
In components it reads
\begin{equation}
{\cal V}^{\mu}_{\alpha}
= J^{\mu}_{\alpha} -2i(\gamma_{\lambda}\theta)_{\alpha} 
\Theta^{\mu \lambda}
+2(\gamma_{5}\theta)_{\alpha} \zeta^{\mu}
+\theta_{\alpha}X^{\mu} 
+{1\over{2}}\, \bar{\theta}\theta f^{\mu}_{\alpha}.
\label{sfcurrentcomponent}
\end{equation}
Here $\zeta^{\mu} = -\epsilon^{\mu\nu} F \partial_{\nu}\phi$, as before,
and
\begin{equation}
X^{\mu}= -\bar{\psi}\,\gamma^{\mu}\gamma^{\nu}\partial_{\nu}\psi
=i\bar{\psi}\, \gamma^{\mu}{\delta S\over{\delta \bar{\psi}}}.
\label{Xmu}
\end{equation}
One may be tempted to set $X^{\mu}=0$, but some care is needed here.
The equation of motion $\delta S/\delta \bar{\psi} =
i\gamma^{\nu}\partial_{\nu}\psi -W''(\phi)\psi=0$ is assumed to hold by
itself. In contrast, the equations of motion multiplied by fields
are potentially singular and, when properly regulated, may not vanish. 
In Fujikawa's path-integral formulation of anomalies~\cite{F,Ftwo} all
known anomalies arise from regularized Jacobian factors which take
precisely the form (fields)$\times$(equations of motion). We shall
therefore avoid using the equations of motion in our analysis and
keep track of the potentially anomalous products such as $X^{\mu}$;
actually, conformal~\cite{ST} and superconformal~\cite{KS} anomalies in
four dimensions were studied along this line earlier.   In the next
section we evaluate such anomalous products by regularizing them in a
supersymmetric way.

The highest term
$f^{\mu}_{\alpha}$ in ${\cal V}^{\mu}_{\alpha}$ has a somewhat long
expression,
\begin{eqnarray}
f^{\mu}_{\alpha} &=& 2\epsilon^{\mu\lambda}
\partial_{\lambda}(F\gamma_{5}\psi)_{\alpha} + r^{\mu}_{\alpha},
\label{highestcurrent}\\
r^{\mu}_{\alpha} &=& (\gamma^{\lambda}\gamma^{\mu} +[\gamma^{\lambda},
\gamma^{\mu}] )_{\alpha\beta} \Big( \psi_{\beta} \partial_{\lambda}
{\delta S\over{\delta F}}  +{\delta S\over{\delta \bar{\psi}_{\beta}}}\,
\partial_{\lambda}\phi \Big) 
\nonumber\\
&& -i(\gamma^{\mu})_{\alpha\beta} \Big( {\delta S\over{\delta
\bar{\psi}_{\beta}}}\, F  + \psi_{\beta} \,{\delta S\over{\delta \phi}}
\Big).
\label{rmu}
\end{eqnarray}
The potentially anomalous products $X^{\mu}$ and $r^{\mu}_{\alpha}$,
vanishing trivially at the classical level, actually continue to vanish at
the quantum level, as shown in the next section; anticipating this
result (and to avoid inessential complications), we shall set
$X^{\mu}=r^{\mu}=0$ in the rest of this section.
Correspondingly,
$f^{0}_{\alpha} = 2\partial_{1}(F\gamma_{5}\psi)_{\alpha}$ is a total
divergence so that the associated conserved spinor
charge vanishes (so long as the field $\psi(x)\rightarrow 0$ as
$x^{1}\rightarrow \pm \infty$),
\begin{equation}
\int_{-\infty}^{\infty} dx^{1}\, f^{0}_{\alpha} = 2
\Big[F(\gamma_{5}\psi)_{\alpha}\Big]^{x^{1}=\infty}_{x^{1}=-\infty} =0.
\end{equation} 
Hence only $Q_{\alpha}$, $P^{\lambda}$ and $Z$ form an
irreducible supermultiplet, yielding a conserved-charge
superfield
\begin{equation}
\int_{-\infty}^{\infty} dx^{1} {\cal V}^{0}_{\alpha}
= Q_{\alpha} -2i(\gamma_{\lambda}\theta)_{\alpha} P^{\lambda}
+2(\gamma_{5}\theta)_{\alpha} Z .
\label{sfcharge}
\end{equation}
Note that, upon supertranslations, Eqs.~(\ref{sfcurrentcomponent}) 
and~(\ref{sfcharge}) correctly lead to the
superalgebras~(\ref{deltaJ}) and~(\ref{superchargealgebra}),
respectively.

Let us examine the conservation law of the supercurrent ${\cal
V}^{\mu}_{\alpha}$.  Let $\partial_{\mu}$ act on it, rewrite 
$(\slash \!\!\! pD)_{\alpha}$ or $(\bar{D}\slash \!\!\! p)_{\alpha}$
in terms of three $D$'s using the formulas in Appendix A, and and
eliminate the expression
$\bar{D}D\Phi$ by use of an identity involving the superfield equation of
motion
\begin{equation}
{\delta S\over{\delta \Phi}} =
 - {1\over{2}}\,\bar{D}D\Phi + W'(\Phi).
\label{dSdPhi}
\end{equation}
The result is
\begin{equation}
\partial_{\mu}{\cal V}^{\mu}_{\alpha} = 
(D_{\alpha}\bar{D}_{\beta}\Phi)\,D_{\beta}{\delta S\over{\delta \Phi}}
-(D_{\alpha}\bar{D}_{\beta}{\delta S\over{\delta
\Phi}})\,D_{\beta}\Phi.
\label{dValpha}
\end{equation}
This conservation law is an identity.
Accordingly, current conservation $\partial_{\mu }{\cal V}^{\mu}_{\alpha}
= 0$ follows from the equation of motion 
$\delta S/\delta\Phi=0$ at the classical level.  
Actually $\partial_{\mu}{\cal V}^{\mu}_{\alpha} = 0$ also persists 
at the quantum level, with the potentially anomalous terms vanishing, as
we shall see later. 

It is enlightening to cast Eq.~(\ref{dValpha}) into a somewhat different
form. Let us multiply it with an arbitrary real spinor superfield
$\Omega_{\alpha}(z)$ (or $\bar{\Omega}=\Omega\gamma^{0}$) 
and integrate over $z=(x^{\mu},\theta_{\alpha})$. 
Making integrations by parts then yields 
\begin{equation}
-\int d^{4}z\, \bar{\Omega}_{\alpha}\partial_{\mu}{\cal V}^{\mu}_{\alpha}
= \int d^{4}z\, {\delta S\over{\delta \Phi}}\, \delta_{\Omega}\Phi,
\label{Ntheorem}
\end{equation}
where $d^{4}z \equiv d^{2}x\, d^{2}\theta$ and
\begin{eqnarray}
\delta_{\Omega}\Phi &=& 
\bar{D}_{\beta}\bar{\Omega}_{\alpha}D_{\alpha}D_{\beta}\Phi
-\bar{D}_{\beta}D_{\alpha}\bar{\Omega}_{\alpha}D_{\beta}\Phi,
\nonumber\\
&=& 
(\bar{D}_{\beta}\bar{D}_{\alpha}\Omega_{\alpha})D_{\beta}\Phi
- 2i(\bar{D}\gamma^{\mu}\Omega)\partial_{\mu}\Phi.
\label{locsusyvariation}
\end{eqnarray}
This shows that the supercurrent ${\cal V}^{\mu}_{\alpha}$ is a Noether
current associated with the superfield variation $\delta_{\Omega}\Phi$
and, in addition, that the right-hand side of Eq.~(\ref{dValpha}) is
essentially a Jacobian factor associated with the field change 
$\Phi + \delta_{\Omega}\Phi$ within Fujikawa's method. 
The $\delta_{\Omega}\Phi$ generalizes translations in superspace to their
local form.  Indeed, with the parametrization 
\begin{equation}
\Omega_{\alpha} = \eta_{\alpha}
- {i\over{4}} (\gamma_{\mu}\theta)_{\alpha}a^{\mu} 
+ {1\over{2}}\, \theta_{\alpha} b 
- {1\over{4}} (\gamma_{5}\theta)_{\alpha}c +
{1\over{2}}\,\bar{\theta}\theta\,\xi_{\alpha} ,
\end{equation}
one finds on the left-hand side of Eq.~(\ref{Ntheorem}) 
the conservation laws
\begin{equation}
-\bar{\xi}_{\alpha}\partial_{\mu}J^{\mu}_{\alpha} 
-a_{\lambda}\partial_{\nu}\Theta^{\nu\lambda}
- b\, \partial_{\mu}X^{\mu} -c\, \partial_{\mu}\zeta^{\mu}
-\bar{\eta}_{\alpha} \partial_{\mu}f^{\mu}_{\alpha}.
\end{equation}
At the same time, the field variation $\delta_{\xi} \Phi$ caused by
$\xi_{\alpha} (x)$ is seen to generalize
supertranslations~(\ref{susycomponent}) into local ones [with 
$\delta F = -i\partial_{\mu}(\bar{\xi}\gamma^{\mu} \psi)$]. Similarly, 
$a^{\mu}(x)$ generates local spacetime translations with the variation  
\begin{equation}
\delta_{a} \Phi = a^{\mu}\partial_{\mu}\phi 
+ \bar{\theta}\Big(a^{\mu} \partial_{\mu}\psi
+ {1\over{2}}\,\psi\, \partial^{\mu}a_{\mu} \Big)
+{1\over{2}}\,\bar{\theta}\theta\,\partial_{\mu}(a^{\mu}F).
\end{equation}
The topological current $\zeta^{\mu}$ is associated with the variation 
\begin{equation}
\delta_{c} \Phi =
-{1\over{2}}\,\bar{\theta}\theta\,
\epsilon^{\mu\lambda}(\partial_{\mu}\phi)\partial_{\lambda}c
\end{equation}
that affects only $F$, as noted in ref.~\cite{FN}, 
while $X^{\mu}(= 0)$ derives from the variation
\begin{equation}
\delta_{b} \Phi
= -i (\bar{\theta}\gamma^{\mu}\psi)\partial_{\mu}b
\end{equation}
that affects $\psi$ alone.
The field variation $\delta_{\eta} \Phi$ caused by
$\eta_{\alpha}(x)$ reveals the form of spinorial
transformations that give rise to the conserved current
$f^{\mu}_{\alpha}$; its explicit form, being unilluminating, is
suppressed here.

The supercurrent  ${\cal V}^{\mu}_{\alpha}$ is also used to generate
superconformal transformations of fields~\cite{FZ}.  
In particular, the conservation laws of superconformal currents are
characterized by the quantity
$(\gamma_{\mu}{\cal V}^{\mu})_{\alpha}$, as seen from the
simplest example
$\partial_{\mu}(\slash \!\!\! x {\cal V}^{\mu})_{\alpha} =
(\gamma_{\mu}{\cal V}^{\mu})_{\alpha}$ for 
$\partial_{\mu}{\cal V}^{\mu}_{\alpha}=0$. 
Rewriting ${\cal V}^{\mu}_{\alpha}$ as
\begin{equation}
{\cal V}^{\mu}_{\alpha}
= (\partial_{\nu}\Phi) (\gamma^{\nu}
\gamma^{\mu})_{\alpha\beta}\,D_{\beta}\Phi
+i{1\over{2}} (\bar{D}D\Phi)\,
(\gamma^{\mu})_{\alpha \beta}\,D_{\beta}\Phi
\end{equation}
and noting $\gamma_{\mu}\gamma^{\lambda}\gamma^{\mu}=0$
yields
$(\gamma_{\mu}{\cal V}^{\mu})_{\alpha} = 
i(\bar{D}D\Phi)\, D_{\alpha}\Phi$, which, by use of
Eq.~(\ref{dSdPhi}), is cast in the form 
\begin{equation}
i(\gamma_{\mu}{\cal V}^{\mu})_{\alpha} + 2 D_{\alpha} W(\Phi) =
2\,{\delta S\over{\delta \Phi}}\, D_{\alpha}\Phi \equiv 
{\cal A}_{\alpha}.
\label{traceidentity}
\end{equation}
This is a supersymmetric version of the trace identity~\cite{CCJ}, 
as seen from the component expression
\begin{eqnarray}
 i(\gamma_{\mu}{\cal V}^{\mu})_{\alpha}
&=&  i(\gamma_{\mu}J^{\mu})_{\alpha}
+2\,\theta_{\alpha}\Theta^{\mu}_{\mu}
+2(\gamma_{5}\theta)_{\alpha}\epsilon_{\mu \lambda}\Theta^{\mu \lambda}
\nonumber\\
&& + 2i(\gamma_{\mu}\theta)_{\alpha}\, \epsilon^{\mu \nu}
\zeta_{\nu} +{1\over{2}}\,
\bar{\theta}\theta\,  (i\gamma_{\mu}f^{\mu})_{\alpha}.
\end{eqnarray}
The currents $i(\gamma_{\mu}J^{\mu})_{\alpha}, \zeta^{\nu}$, etc., thus
form a supermultiplet with  $\Theta^{\mu}_{\mu}$.
Normally the trace identity suffers from an anomaly in
theories with ultraviolet divergences.  
Indeed, ${\cal A}_{\alpha} =2 (\delta S/\delta\Phi)\, D_{\alpha}\Phi$
turns out nonvanishing at the one-loop level, and this in turn forces
a quantum modification of the topological current $\zeta^{\nu}$, as shown
in the next section.

One may, using a general real spinor superfield $\omega_{\alpha}(z)$, 
cast Eq.~(\ref{traceidentity}) in the form
\begin{equation}
\int d^{4}z\, \bar{\omega}_{\alpha} \Big\{i(\gamma_{\mu}{\cal
V}^{\mu})_{\alpha} + 2D_{\alpha}W(\Phi) \Big\} = 
\int d^{4}z\, {\delta S\over{\delta \Phi}}\, \delta_{\omega}\Phi.
\end{equation}
This shows again that the anomalous term ${\cal A}_{\alpha}$ is
essentially a Jacobian factor associated with the local field variation
\begin{equation}
\delta_{\omega}\Phi = 2\bar{\omega}_{\alpha}D_{\alpha}\Phi.
\end{equation}
Similarly, the conservation law for $i(\slash \!\!\! x
{\cal V}^{\mu})_{\alpha}$ is seen to follow from the field variation
\begin{equation}
\delta\Phi = \delta_{\Omega}\Phi
-2\,\bar{\omega}_{\alpha}D_{\alpha}\Phi,
\label{sctransf}
\end{equation}
where $\delta_{\Omega}\Phi$ stands for the superfield
variation~(\ref{locsusyvariation}) with 
$\Omega_{\alpha} = -i(\slash \!\!\! x \omega)_{\alpha}$.
This $\delta\Phi$ correctly involves, e.g., for $\omega_{\alpha}
\rightarrow {1\over{2}}\, \theta_{\alpha} b$ (with constant $b$),
the field transformation laws under dilatation,
$\delta \phi_{i} = 2b \{x^{\mu}\partial_{\mu}\phi_{i} 
+ (d_{i}/2)(\partial_{\mu}x^{\mu})\phi_{i}\}$ with $d_{i}=(0,1/2,1)$ 
for $\phi_{i}=(\phi, \psi, F)$.  Equation~(\ref{sctransf}) isolates 
from $\delta\Phi$ an anomaly-free variation $\delta_{\Omega}\Phi$.
This shows that the anomalous term  ${\cal A}_{\alpha}$, 
governed by the $\delta_{\omega}\Phi$ variation, essentially constitutes
the superconformal anomaly.

\section{Anomalies}

In this section we evaluate potentially anomalous equation-of-motion-like
terms and discuss their consequences.
For operator calculus in superspace it is useful to define  
the eigenstate  of the coordinates 
$z=(x^{\mu},\theta_{\alpha})$ by $|z\rangle =|x\rangle |\theta)$, with
$\theta_{\alpha} |\theta') =\theta'_{\alpha} |\theta')$ and
normalization $(\theta'|\theta'') 
={1\over{2}}\, (\bar{\theta}' -\bar{\theta}'') (\theta' -\theta'') 
\equiv \delta (\theta' -\theta'')$.
The covariant derivative $D_{\alpha}$ is thereby written as an
antisymmetric matrix
$\langle z|D_{\alpha}|z'\rangle  = D_{\alpha}\delta (\theta
-\theta')\delta^{2} (x -x') =-\langle z'|D_{\alpha}|z\rangle$.
A direct calculation shows that
\begin{equation}
(\theta|\bar{D}D |\theta')\equiv \bar{D}D\delta (\theta -\theta')
= -2 \exp[\bar{\theta}{\slash\!\!\! p} \theta'],
\label{DbarD}
\end{equation}
where $p_{\mu}=i\partial_{\mu}$.
Thus, for the $\theta$-diagonal element
\begin{equation}
(\theta|-{1\over{2}}\bar{D}D |\theta)=1,
\end{equation}
whereas odd numbers of $D$'s have vanishing diagonal elements
$(\theta|D_{\alpha}|\theta)=(\theta|D_{\alpha}\bar{D}D|\theta)=0$, etc.,
as verified easily.  This feature of $D$'s plays an important role in
superspace operator calculus.

The potentially anomalous products of our concern have the
form
\begin{equation}
\{\Lambda \Phi (z) \}\, \Xi {\delta S\over{\delta \Phi (z)}} ,
\label{XiLambda}
\end{equation}
where $\Lambda$ and $\Xi$ may involve operators $D_{\alpha}$ and
$\partial_{\mu}$.
To evaluate them we use the background-field method~\cite{CW} and
decompose 
$\Phi$ into the classical field
$\Phi_{\rm c}$ and the quantum fluctuation $\chi$,
$\Phi (z) = \Phi_{\rm c}(z) + \chi(z)$.
The quantum fluctuation at the one-loop level is governed by the action
\begin{eqnarray}
 S_{\chi}&=& \int d^{4}z\,  
{1\over{2}}\, \chi(z) \,{\cal D}\, \chi (z), 
\label{Schi}\\
{\cal D}&=&-{1\over{2}}\bar{D}D + W''(\Phi_{\rm c}).
\end{eqnarray}
The associated $\chi$ propagator is given by $i/{\cal D}$,
which we regularize as 
\begin{eqnarray}
&&\langle \chi (z)\,\chi (z')\rangle^{\rm reg} = \langle z|
{i\over{\cal D}}\, \Gamma |z'\rangle,
\label{regpropagator} \\
&&\Gamma = e^{\tau {\cal D}^{2}}, 
\end{eqnarray}
with $\tau \rightarrow 0_{+}$ in $\Gamma$ at the very end.
This choice $\Gamma = e^{\tau {\cal D}^{2}}$ of the ultraviolet regulator
is manifestly supersymmetric and is a natural one~\cite{FN,ST} that
controls quantum fluctuations within the background-field method.

To evaluate Eq.~(\ref{XiLambda}) at the one-loop level 
we set $\Lambda \Phi \rightarrow \Lambda \chi$ and 
$\Xi (\delta S/\delta \Phi) 
\rightarrow  \Xi (\delta S_{\chi}/\delta \chi)$, 
consider as the singular part the expectation value, and substitute the
regularized $\chi$ propagator~(\ref{regpropagator}).
The result is
\begin{equation}
\langle  \{\Lambda \Phi(z)\} \, \Xi  {\delta S\over{\delta \Phi
(z)}} \rangle = i\, \langle z|\, \Lambda \Gamma \Xi^{T}|z\rangle 
= \pm i\, \langle z|\, \Xi \Gamma \Lambda^{T}|z\rangle, 
\label{regJacobian}
\end{equation}
where the minus sign $-$ applies only when both $\Lambda$ and $\Xi$ are
Grassmann-odd.
Here transposition $\Lambda^{T}$ is defined by
$\langle z|\Lambda^{T}|z'\rangle \equiv \langle z'|\Lambda|z\rangle$;
in particular, $(D_{\alpha})^{T}= -D_{\alpha}$ and $\Gamma^{T} = \Gamma$.
It is readily seen that 
the Leibnitz rule applies to the regularized products, 
\begin{equation}
(\partial_{\mu}\Lambda \Phi)\, \Xi {\delta S\over{\delta \Phi}}
+(\Lambda \Phi)\, \partial_{\mu}\Xi {\delta S\over{\delta \Phi}}
= \partial_{\mu}\Big[ (\Lambda \Phi)\,  \Xi{\delta S\over{\delta
\Phi}} \Big]; 
\end{equation}
an analogous formula holds with $D_{\alpha}$ as well.

From Eq.~(\ref{regJacobian}) follows a key formula for regularized
products,
\begin{equation}
(\Lambda \Phi)\, \Xi {\delta S\over{\delta \Phi}}
= \pm\, (\Xi \Phi)\, \Lambda {\delta S\over{\delta \Phi}}, 
\label{formulaJacobian}
\end{equation}
where the minus sign applies only when both $\Lambda$ and $\Xi$ are
Grassmann-odd.  
An immediate consequence of this formula is the
conservation of the supercurrent $\partial_{\mu}{\cal V}^{\mu}_{\alpha}=0$; 
indeed, setting
$\Lambda
\rightarrow D_{\alpha}\bar{D}_{\beta}$ and 
 $\Xi \rightarrow D_{\beta}$ in Eq.~(\ref{formulaJacobian})
implies the vanishing of  the potentially anomalous term in
Eq.~(\ref{dValpha}), irrespective of the choice of
$\Gamma$.

Similarly one can show the vanishing of the potentially anomalous products
$X^{\mu}$ and $r^{\mu}_{\alpha}$ [in Eqs.~(\ref{Xmu}) and~(\ref{rmu})]
without  any direct calculation. To this end, note first the formula  
\begin{equation}
{\delta S\over{\delta \Phi}}
= {\delta S\over{\delta F}} - \bar{\theta}\,
 {\delta S\over{\delta \bar{\psi}}} +{1\over{2}}\,\bar{\theta}\theta\,
{\delta S\over{\delta \phi}} ,
\end{equation}
with which one can relate superfield expressions to component-field
expressions.
For $X^{\mu}=i\bar{\psi}\, \gamma^{\mu}(\delta S/\delta\bar{\psi})$ 
one may start with the superfield expression
$(D_{\alpha} \Phi)\, D_{\beta} (\delta S/\delta \Phi)$, 
which, in view of the formula~(\ref{formulaJacobian}), is seen to be 
antisymmetric in $(\alpha,\beta)$, or proportional to
$(\gamma^{0})_{\alpha\beta}$; hence
\begin{equation}
(\bar{D}_{\alpha} \Phi)\, D_{\beta} {\delta S\over{\delta \Phi}}
\propto \delta_{\alpha \beta}, 
\label{DaDb}
\end{equation}
which, in component fields, implies that 
\begin{eqnarray}
&&X^{\mu}=i\bar{\psi}\,\gamma^{\mu}(\delta S/\delta\bar{\psi}) =0,\ \ 
\bar{\psi}\, \gamma_{5}(\delta S/\delta\bar{\psi}) =0,\nonumber\\
&&\bar{\psi}\,(\delta S/\delta\bar{\psi}) \not=0, {\rm etc.}
\end{eqnarray}
For $r^{\mu}_{\alpha}$ note first the regularized superfield
relation following from Eq.~(\ref{formulaJacobian}):
\begin{equation}
{1\over{2}}(\bar{D}D\Phi)\, D_{\alpha} {\delta S\over{\delta \Phi}} 
- (D_{\alpha}\Phi)\, {1\over{2}}\bar{D}D{\delta S\over{\delta\Phi}}
=0,
\end{equation}
which shows the vanishing of the combination 
$F\, (\delta S/\delta \bar{\psi}) + \psi \,(\delta S/\delta \phi)$ 
in $r^{\mu}_{\alpha}$.
The vanishing of the remaining combination 
 $\psi\, \partial_{\lambda} (\delta S/\delta F) 
+(\delta S/\delta \bar{\psi})\, \partial_{\lambda}\phi$
also follows from an analogous equation with ${1\over{2}}\bar{D}D$
replaced by $\partial_{\lambda}$ in the above. 
Hence $r^{\mu}_{\alpha}=0$.

On the other hand, the supersymmetric trace identity~(\ref{traceidentity})
does acquire a quantum modification
\begin{equation}
{\cal A}_{\alpha}  \equiv  2\,{\delta S\over{\delta \Phi}}\,
D_{\alpha}\Phi = iD_{\alpha} \langle z|\Gamma\,|z\rangle,
\label{Aalpha}
\end{equation}
where the last expression follows from 
${\cal A}_{\alpha}  = -2i \langle z| \Gamma D_{\alpha}|z\rangle 
= 2i \langle z|D_{\alpha}\Gamma\,|z\rangle
= i\langle z|[D_{\alpha},\Gamma]\,|z\rangle$.
A direct calculation of the relevant heat-kernel is outlined 
in Appendix B.  The result is
\begin{equation}
{\cal A}_{\alpha}  
= - {1\over{2\pi}}\, D_{\alpha}W''(\Phi_{\rm c}).
\end{equation}
One may now set $\Phi_{\rm c} \rightarrow \Phi$ 
in ${\cal A}_{\alpha}$ to obtain the corresponding operator
expression. 
For the trace identity the net effect of the anomaly is to replace 
$W(\Phi)$ by  $ W(\phi) +(1/4\pi)\, W''(\Phi)$ on the right-hand side:
\begin{equation}
i(\gamma_{\mu}{\cal V}^{\mu})_{\alpha} 
=-2D_{\alpha}\Big\{ W(\Phi)+ {1\over{4\pi}}\, W''(\Phi) \Big\}.
\end{equation}
For component fields this implies that
\begin{eqnarray}
&&i(\gamma_{\mu}J^{\mu})_{\alpha}= 2F \psi_{\alpha} = 
- 2 \Big( W' +{1\over{4\pi}}\, W'''\Big) \psi_{\alpha}, \nonumber\\
&&\Theta^{\mu}_{\mu}= F^{2}+{i\over{2}}\, 
\bar{\psi} \gamma^{\mu}\partial_{\mu}\psi \nonumber\\
&&\ \ \ \ \,
=\!  -F \Big( W' \!+\!{1\over{4\pi}}\, W'''\Big) \!+\!
{1\over{2}} \Big( W'' \!+\!{1\over{4\pi}}\, W''''\Big) \bar{\psi}\psi,
\nonumber\\
&&\epsilon_{\mu \lambda}\Theta^{\mu \lambda}
= {i\over{2}}\, \bar{\psi} \gamma_{5}\gamma^{\mu}
\partial_{\mu}\psi = 0,\nonumber\\
&&\zeta^{\mu}= -\epsilon^{\mu\nu} F\partial_{\nu}\phi 
= \epsilon^{\mu \nu}
\partial_{\nu} \Big( W +{1\over{4\pi}}\, W''\Big), {\rm etc.;}
\label{sconfanomaly}
\end{eqnarray}
these component-field expressions agree with the result of
ref.~\cite{SVV}.

It is seen from Eq.~(\ref{sconfanomaly}) that the role of the
auxiliary field $F$ in composite operators is modified at the quantum
level: $F$ multiplied by
$\psi_{\alpha}$ or
$\partial_{\nu}\phi$ acts like $-\{ W'(\phi) +(1/4\pi)\, W'''(\phi)\}$
rather than $-W'(\phi)$.  Note that such modifications take place in
the supercurrent~(\ref{sfcurrentcomponent}) itself, e.g., 
in the $F \psi_{\beta}$ terms of $J^{\mu}_{\alpha}$ and
$f^{\mu}_{\alpha}$, and in the $F\partial_{\nu}\phi$ term of
$\zeta^{\mu}$ (if one were to eliminate $F$ from these). 
The naive canonical expressions for $\zeta^{\mu}$ and $Z$ in
Eqs.~(\ref{zetanaive}) and~(\ref{Znaive}) are not valid any more;
$\zeta^{\mu}$ in Eq.~(\ref{sconfanomaly}) leads to the modified central
charge~\cite{SVV}
\begin{equation}
Z = \int dx^{1} \zeta^{0}
=\Big[ W(\phi) +
{1\over{4\pi}}W''(\phi)\Big]^{x^{1}=\infty}_{x^{1}=-\infty}.
\label{Zanom}
\end{equation}

It is enlightening to look into some other definitions of the superfield
supercurrent.
Consider a supercurrent with $F$ replaced by $-W'$:
\begin{equation}
\hat{J}^{\mu}_{\alpha} \equiv  (\gamma^{\nu}\gamma^{\mu}\psi)_{\alpha}\,
\partial_{\nu}\phi +i(\gamma^{\mu}\psi)_{\alpha}\,W'(\phi),
\end{equation}
which differs from $J^{\mu}_{\alpha}$ in Eq.~(\ref{Jmualpha}) by 
$i(\gamma^{\mu}\psi)_{\alpha}\,
(\delta S/\delta F)$. This is a supercurrent one would obtain in
the on-shell realization of supersymmetry. 
The superfield generalization of this current is
\begin{equation}
\hat{{\cal V}}^{\mu}_{\alpha} =
(\partial_{\nu}\Phi) (\gamma^{\nu}
\gamma^{\mu})_{\alpha\beta}\,D_{\beta}\Phi
+iW'(\Phi)\, (\gamma^{\mu}D)_{\alpha}\Phi,
\label{Vhat}
\end{equation}
which, when expanded in $\theta$ as in Eq.~(\ref{sfcurrentcomponent}),
involves the components
\begin{eqnarray}
\hat{\Theta}^{\mu \lambda}&=&
{i\over{4}}\, \bar{\psi} (\gamma^{\mu} \partial^{\lambda }
+\gamma^{\lambda} \partial^{\mu})\psi+
\partial^{\mu}\phi\partial^{\lambda}\phi \nonumber\\
&&
- {1\over{2}}\, g^{\mu \lambda}\Big\{(\partial\phi)^{2} +FW' 
+ {1\over{2}}\,  \bar{\psi}(i\gamma\! \cdot\!\partial -W'')\psi \Big\},
\nonumber\\
 \hat{\zeta}^{\mu}&=& (1/2)\,
\epsilon^{\mu\nu}(\partial_{\nu}W - F\partial_{\nu}\phi -X_{\nu}),
\nonumber\\
\hat{X}^{\mu}&=&
F\partial^{\mu}\phi + \partial^{\mu}W + (1/2)\, X^{\mu},\ {\rm etc.}
\end{eqnarray}
Here we find the symmetric energy-momentum
tensor; remember, however, that $\hat{\Theta}^{\mu \lambda}$ and 
$\Theta^{\mu \lambda}$ are different since 
$\hat{\Theta}^{\mu}_{\mu} \not= \Theta^{\mu}_{\mu}$ at the quantum level.
Also, $\hat{\zeta}^{\mu}=\zeta^{\mu} + (1/8\pi) \epsilon^{\mu \nu}
\partial_{\nu}W'' \not= \zeta^{\mu}$ and $\hat{X}^{\mu}= -(1/4\pi)
\partial^{\mu}W'' \not=0$.

The current $\hat{{\cal V}}^{\mu}_{\alpha}$ differs from 
${\cal V}^{\mu}_{\alpha}$ by an anomalous field product, actually by the
anomaly ${1\over{2}} (i\gamma^{\mu}{\cal A})_{\alpha}$,
\begin{equation}
\hat{{\cal V}}^{\mu}_{\alpha} = {\cal V}^{\mu}_{\alpha}
+ i(\gamma^{\mu}D\Phi)_{\alpha}(\delta S/\delta \Phi).
\end{equation}
As a result, with $\hat{{\cal V}}^{\mu}_{\alpha}$, supersymmetry is
apparently spoiled while the trace identity looks normal,
\begin{eqnarray}
&&\partial_{\mu}\hat{{\cal V}}^{\mu}_{\alpha}
= -(i/4\pi)\, \partial_{\mu}[\gamma^{\mu} D_{\alpha}W''(\Phi)]
\not=0, \\
&&(\gamma_{\mu}\hat{{\cal V}}^{\mu})_{\alpha} = 2i D_{\alpha}W(\Phi).
\end{eqnarray}
Here we encounter trading of the anomaly between
two different conservation laws, a phenomenon observed also in
ref.~\cite{FN} and familiar from nonsupersymmetric theories~\cite{CCJ}. 
Since the trace identity is anomalous in theories with divergences, 
it is appropriate to recover supersymmetry, 
which is achieved by passing from 
$\hat{{\cal V}}^{\mu}_{\alpha}$ to ${\cal V}^{\mu}_{\alpha}$ through
current redefinition. In this sense, 
${\cal V}^{\mu}_{\alpha}$ is a more natural off-shell definition 
of the supercurrent.

We have so far studied potentially anomalous products in the one-loop
approximation. The basic formula~(\ref{regJacobian}) with 
$\Gamma = e^{\tau {\cal D}^{2}}$ itself is an expression valid at the
one-loop level.  Fortunately it is generalized to higher loops if one
notes that the cutoff $\Gamma$ there derives from the expression
\begin{equation}
\langle z | \Gamma |z' \rangle 
= -i{\cal D}_{z}\langle \chi (z) \chi(z') \rangle^{\rm reg};
\end{equation}
that is, one may extract the relevant cutoff $\Gamma$ from the
$\chi$ propagator calculated to any desired loop levels using the
regularized zero$th$-order $\chi$ propagator~(\ref{regJacobian}).

Actually it is not necessary to study higher-loop corrections
in the present case. 
The central charge anomaly or the supersymmetric
anomaly ${\cal A}_{\alpha}$ is one-loop exact and there are no
higher-loop corrections, as pointed out in ref.~\cite{SVV}. This
follows from a dimensional analysis if one notes that the
anomalies in local field products come from short distances,
as they should: 
Note that ${\cal A}_{\alpha}= 2\,(\delta S/\delta \Phi)\,
D_{\alpha}\Phi$ has dimension 3/2 in units of mass while higher-loop
corrections, calculated in a fashion outlined above, inevitably supply at
least two powers of 
$W''[\Phi_{\rm c}], W'''[\Phi_{\rm c}],\cdots$, each having dimension
one; this implies the one-loop exactness of the anomaly.  (No inverse
powers of $W$ are allowed as long as the anomaly is short-distance
dominated.)

\section{Superspace effective action}

In this section we verify the anomaly by a direct calculation of the 
effective action in superspace.
Setting $\Phi (z) \rightarrow \Phi_{c}(z)$ in the action~(\ref{sfaction})
yields the classical action $S[\Phi_{c}]$.
The one-loop effective action $\Gamma_{1}[\Phi_{c}]$ is a functional
of $M = W''(\Phi_{c})$, as seen from $S_{\chi}$ in Eq.~(\ref{Schi}), and
is most efficiently calculated from its derivative
$\delta \Gamma_{1}[\Phi_{\rm c}]/\delta \Phi_{\rm c}(z)$,
which is related to the propagator 
$\langle \chi (z)\,  \chi (z')\rangle =  
i\langle z| ( -{1\over{2}}\, \bar{D}D + M)^{-1}|z'\rangle$,
\begin{equation}
\delta \Gamma_{1}[\Phi_{\rm c}]/\delta \Phi_{\rm c}(z)
= {1\over{2}}\,
 M' \langle  \chi (z)\,  \chi (z) \rangle, 
\label{dGammadphi}
\end{equation}
where  $M' \equiv dM/d\Phi_{\rm c} = W'''(\Phi_{\rm c})$.
One can evaluate the $\chi$ propagator by expanding it in powers of
$D_{\alpha}$ acting on $M$; see Appendix C for details. 
To $O(D^{2})$  the result is 
\begin{equation}
\langle \chi (z)\,  \chi (z)\rangle 
= {1\over{4\pi}}\, \Big[ \ln {\Lambda^{2}\over{M^{2}}} 
- {\bar{D}DM \over{2 M^{2}}}+ {(\bar{D}_{\alpha}M) (D_{\alpha}M)
\over{2 M^{3}}}
\Big],
\label{chipropagator}
\end{equation}
where $\Lambda$ is an ultraviolet cutoff; $\Lambda^{2} =
e^{-\gamma}/\tau$ if one adopts the regularized
propagator~(\ref{regpropagator}). Substituting this into
Eq.~(\ref{dGammadphi}) and  integrating with respect to $\Phi_{\rm c}$
then yields the one-loop effective action to $O(D^{2})$,
\begin{equation}
\Gamma_{1}[\Phi_{\rm c}]
=  \int d^{4}z\,  {1\over{8\pi}}\, \Big[  
M\Big( \ln {\Lambda^{2}\over{M^{2}}} + 2 \Big)  
+ {(\bar{D}M) (DM) \over{4 M^{2}}} \Big].
\end{equation}

Let us here review briefly how such quantum corrections could relate to
the central charge anomaly~\cite{WO,SVV}.
As we have seen, supertranslations have no quantum anomaly.  This implies
that the supermultiplet structure of $Q_{\alpha}, P^{\mu}$ and $Z$ in
Eq.~(\ref{sfcharge}), or equivalently the supercharge
algebra~(\ref{superchargealgebra}), is maintained exactly.  
For static kink states the superalgebra~(\ref{superchargealgebra}) is
split into two algebras
\begin{equation}
(Q_{1})^{2} = P^{0} +Z, (Q_{2})^{2} = P^{0} - Z.
\end{equation}
The classical kink solution~(\ref{WBkink}), obeying the 
first-order equation $(\partial/\partial x^{1}) \phi = W'(\phi)$, is
BPS saturated in the sense that it is inert under supertranslations
generated by the supercharge $Q_{2}$, or equivalently, $Q_{2}$ annihilates
the kink state~\cite{WO},
\begin{equation}
Q_{2} |{\rm kink}\rangle  = (P^{0} - Z)|{\rm kink}\rangle =0.
\label{BPSsaturation}
\end{equation}
The supercharge $Q_{1}$, on the other hand, acts nontrivially, leading to
two degenerate kink states, one bosonic and one fermionic.  
The BPS saturation~(\ref{BPSsaturation}), 
once established classically, persists~\cite{W,SVV,LSV} at the
quantum level (at least in perturbation theory), and 
the kink mass is related to the central charge
$\langle {\rm kink}|Z| {\rm kink}\rangle$ exactly.

Let us verify the above formal discussion using the effective action
$S[\Phi_{\rm c}] + \Gamma_{1}[\Phi_{\rm c}]$.
With only the bosonic components $\phi_{\rm c}(x)$ and $F_{\rm c}(x)$ of
$\Phi_{\rm c}(z)$ retained, $\Gamma_{1}[\Phi_{\rm c}]$ reads
\begin{equation}
\Gamma_{1}
= \int d^{2}x\,  {1\over{8\pi}}\, \Big[  
F_{\rm c} M'_{\phi} \ln {\Lambda^{2}\over{M_{\phi}^{2}}}   
+ {1\over{2}}\Big({M'_{\phi}\over{M_{\phi}}}\Big)^{2}
\{ (\partial_{\mu}\phi_{\rm c})^{2} + F_{\rm c}^{2} \} \Big],
\end{equation}
where $M_{\phi}\equiv W''(\phi_{\rm c})$.
Accordingly, the static kink is now governed by the
Lagrangian
\begin{equation}
 {\cal L}_{\rm stat}
=  -{1\over{2}}\,\Big(\sqrt{\alpha}\, (\partial_{1}\phi_{\rm c})
\mp {1\over{\sqrt{\alpha}}}\, W'_{\rm eff}\Big)^{2} \mp
W'_{\rm eff}\partial_{1}\phi_{\rm c},
\end{equation}
where 
\begin{equation}
\alpha (\phi_{\rm c})= 1 + 
{1\over{8\pi}}\,\Big({M'_{\phi}\over{M_{\phi}}}\Big)^{2}, \ \ 
W'_{\rm eff}(\phi_{\rm c}) = W'(\phi_{\rm c}) + { M'_{\phi}\over{8\pi}} 
\ln {\Lambda^{2}\over{M_{\phi}^{2}}} .
\end{equation}
This leads to the BPS equation for the kink,
\begin{equation}
\partial_{1}\phi_{\rm c}
= -F_{\rm c}= (1/\alpha)\, W'_{\rm eff}(\phi_{\rm c}),
\label{BPSeq}
\end{equation}
with the asymptotic values of $\phi_{\rm c}$ at $x^{1}=\pm \infty$ now
determined by $W'_{\rm eff}(\phi_{\rm c}) = 0$ and the kink mass related
to the central charge 
$Z_{\rm eff} = \int dx\,  W'_{\rm eff}\partial_{1}\phi_{\rm c}$, 
\begin{equation}
m^{\rm kink} = Z_{\rm eff}
= 2W_{\rm eff}(\phi_{\rm c})|_{x^{1}=\infty},
\end{equation}
where
\begin{equation}
W_{\rm eff}(\phi_{\rm c}) =  W(\phi_{\rm c}) + 
{W''\over{8\pi}}\,   
\Big\{ \ln {\Lambda^{2}\over{ (W'')^{2}}} + 2 \Big\}.
\label{Zeff}
\end{equation}

Note here that in the background-field method the superpotential
$W(\Phi)$ acquires a one-loop correction of the form,
\begin{equation}
\langle W(\Phi) \rangle = W(\Phi_{\rm c}) 
+ {1\over{2}}\, W''(\Phi_{\rm c}) \langle \chi (z) \chi(z) \rangle 
+ \cdots.
\end{equation}
Using Eq.~(\ref{chipropagator}) and comparing $\langle W(\Phi) \rangle$
with $W_{\rm eff}(\phi_{\rm c})$, one learns that 
the $(W''/8\pi)\, \ln [\Lambda^{2}/(W'')^{2}]$ term 
(+ a term $\propto W' \approx 0$) in $W_{\rm eff}$ is
essentially a one-loop contribution from $W(\Phi)$ and that the central
charge  $Z_{\rm eff}$ correctly involves the anomaly term
$(1/4\pi)\, W''(\phi)$ of Eq.~(\ref{Zanom}).

On the other hand, the same line of argument shows that the equation of
motion for $\Phi_{\rm c}$ following from 
$S[\Phi_{\rm c}] +\Gamma_{1}[\Phi_{\rm c}]$ is neatly summarized by 
\begin{equation}
 -{1\over{2}}\, \bar{D}D\Phi_{\rm c} + \langle W'(\Phi) \rangle =0.
\label{eqofmotion}
\end{equation}
This verifies that the operator equation of motion $\delta S/\delta \Phi
=0$ by itself is normal, yielding no anomaly.
The effective action thus correctly embodies both the equation of
motion and the quantum anomaly. 

The ultraviolet cutoff $\Lambda^{2}$ which we have kept so far
can be removed by renormalization.
Let us consider the Wess-Zumino model with the
superpotential~(\ref{WZmodel}). It is superrenormalizable and only mass
renormalization is needed. Let $m_{\rm r}$ be a finite mass scale and set
$m^{2}=m_{\rm r}^{2}  + \delta m^{2}$ in the effective action
$S[\Phi_{\rm c}] + \Gamma_{1}[\Phi_{\rm c}]$.
A convenient choice for the mass counterterm is 
\begin{equation}
\delta m^{2} = (\lambda^{2}/\pi) \ln (\Lambda^{2}/m_{\rm r}^{2}),
\end{equation}
the net effect of which is to 
replace the cutoff $\Lambda^{2}$ by $m_{\rm r}^{2}$ 
[and $m^{2}$ by $m_{\rm r}^{2}$] in 
$S[\Phi_{\rm c}] + \Gamma_{1}[\Phi_{\rm c}]$  
or in $W_{\rm eff}(\phi_{\rm c})$ of Eq.~(\ref{Zeff}). 
Minimizing $W_{\rm eff}(\phi_{\rm c})$ then gives the kink mass
\begin{eqnarray}
m^{\rm kink} = 2W_{\rm eff}(\phi_{\rm c})|_{x^{1}=\infty}
= {m_{\rm r}^{3}\over{6\lambda^{2}}} -{m_{\rm r}\over{2\pi}}
\end{eqnarray}
at $\phi_{\rm c}(x^{1}=\infty) = m_{\rm r}/(2\lambda)$, in agreement
with earlier results~\cite{NSNR,SVV}. 

For the sine-Gordon model with $W(\Phi)= m v^{2} \sin (\Phi/v)$ one may
set $m=m_{\rm r}  + \delta m$ and choose
\begin{equation}
\delta m = (m_{\rm r}/8\pi v^{2}) \ln (\Lambda^{2}/m_{\rm r}^{2}).
\end{equation}
This leads to $\phi_{\rm c}(x^{1}=\infty) = (\pi/2)\, v$ and the soliton
mass
\begin{eqnarray}
m^{\rm sol} = 2W_{\rm eff}(\phi_{\rm c})|_{x^{1}=\infty}
= 2m_{\rm r} v^{2} -(m_{\rm r}/2\pi).
\end{eqnarray}

Finally some remark are in order. In the above we have used
the one-loop effective action $\Gamma_{1}[\Phi_{\rm c}]$ to $O(D^{2})$.
Corrections of higher powers of $D_{\alpha}$ lead to more than two
derivatives $\partial_{1}$ acting on $\phi_{\rm c}$ and hardly affect
the asymptotic behavior of $\phi_{\rm c}$ 
for  $x^{1} \rightarrow \pm \infty$, thus leaving the above discussion on
BPS saturation intact.   It is enlightening to discuss the saturation
using superfields. In terms of superfields the BPS
saturation~(\ref{BPSsaturation}) implies that the bosonic-kink superfield
has the form~\cite{SVV}
\begin{equation}
\Phi_{\rm c}(z)
= \phi_{\rm c} \Big( x^{1}  -{1\over{2}}\,\bar{\theta}\theta \Big).
\end{equation}
Substituting this into the equation of
motion~(\ref{eqofmotion}) (and noting that $D_{1}\Phi_{\rm c}=0$ and
$\bar{D}D\Phi_{\rm c}= 2\partial_{1}\phi_{\rm c}$) yields precisely the
BPS equation~(\ref{BPSeq}); here, unlike the component-field
equations of motion, the superfield equation of motion has directly
turned into the BPS equation.  This offers direct verification that the
BPS saturation continues beyond the classical level.

\section{Summary and discussion}

In the present paper we have studied, within a manifestly supersymmetric
setting of the superfield formalism, anomalies and quantum corrections to
solitons in two-dimensional theories with $N=1$ supersymmetry. 
Extensive use is made of the superfield supercurrent to study the
structure of supersymmetry and related superconformal
symmetry in the presence of solitons, and to make explicit the
supermultiplet structures of various symmetry currents and their
associated anomalies.

Possible anomalies in the conservation laws of the supercurrent
and associated currents are calculated from potentially
anomalous products of the form  (fields)$\times$(equations of motion).
Interestingly, such potentially anomalous products appear not only in the
current conservation laws but also in the superfield supercurrent itself.
This does not imply that the current itself is anomalous.

It is perhaps worthwhile to make this point clear and to summarize how
the central charge anomaly arises.
In the algebraic (i.e., off-shell)  realization of supersymmetry one
inevitably handles the auxiliary field $F$, which in the on-shell
realization is eliminated in favor of the basic bosonic field $\phi$.
The superfield ${\cal V}^{\mu}_{\alpha}$ in Eq.~(\ref{sfcurrent}) is
an algebraically natural definition of the supercurrent
(because it is conserved, as we have seen).
In passing to an on-shell expression for the supercurrent one has to 
replace $F$ by $-W'(\phi) + \delta S/\delta F$.  
The field products involving $F$ thereby reveal portions
that lead to possible deviations from classical expressions. 
Indeed, the relevant superfield product is
\begin{equation}
-(\bar{D}D\Phi)\,  D_{\alpha}\Phi = -D_{\alpha}W(\Phi) 
+ {\cal A}_{\alpha}
\end{equation}
in Eq.~(\ref{traceidentity}), owing to which the product $F\psi_{\beta}$
in $J^{\mu}_{\alpha}$ and $f^{\mu}_{\alpha}$ has turned into 
$- \{W' + (1/4\pi) W''\}\psi_{\beta}$, and the product
$F\partial_{\mu}\phi$ in $\zeta^{\mu}$ into 
$- \{W' + (1/4\pi) W''\}\partial_{\mu }\phi$. 
The auxiliary field $F$ thus changes its role in each composite
operator~\cite{fn}. This is one way to understand the emergence of
anomalies. Actually, since the relevant portion of such changes in the
role played by $F$ is associated with superconformal transformations,
as we have seen in Sec.~III, it may be said that 
the central charge anomaly is part of the superconformal
anomaly.
It is important to note here that the supercurrent, untouched
in the off-shell realization, is only apparently modified in passing to
the on-shell realization of supersymmetry; in this way the supercurrent
${\cal V}^{\mu}_{\alpha}$ is capable of accommodating the superconformal
anomaly without spoiling supersymmetry.

If, on the other hand, one avoids the use of $F$, thus necessarily
relying on the on-shell realization of supersymmetry, the superfield
supercurrent [e.g., $\hat{\cal V}^{\mu}_{\alpha}$ in Eq.~(\ref{Vhat})] is
no longer able to admit the effect of the anomaly and supersymmetry gets
apparently spoiled; then redefinition of currents is needed to recover
supersymmetry.

The combined use of the background-field method and Fujikawa's
path-integral formulation of anomalies is crucial in our discussion
of supersymmetric anomalies; remember in this respect that the way the
anomalies arise depends critically on the regularization method one
uses.   The superspace background-field method provides a neat means for
supersymmetric calculations [with a very natural regularization
prescription~(\ref{regpropagator})] and is particularly suited for the
discussion of solitons and other topological excitations.   As we have
seen in Sec.~V, it would be advantageous, especially when direct
calculations of the soliton mass and/or central charge involve some
subtleties, to first study the effective action which reveals how the
soliton equation and the associated boundary condition, as well as the
central charge, are modified at the quantum level.

\acknowledgments

The author wishes to thank K. Fujikawa and P. van Nieuwenhuizen for
calling his attention to central charge anomalies in supersymmetric
theories.  This work was supported in part by a Grant-in-Aid for
Scientific Research from the Ministry of Education of Japan, Science and
Culture (Grant No. 14540261).

\appendix

\section{Some useful formulas}

In this appendix we summarize some formulas involving
spinor derivatives $D_{\alpha}$. 
Let us first write, using the algebra
$\{\bar{D}_{\alpha}, D_{\beta}\} = 2(\slash\!\!\! p)_{\beta \alpha}$
with
$p_{\mu}=i\partial_{\mu}$,  products of two $D$'s in the form 
\begin{eqnarray}
&&\bar{D}_{\alpha}D_{\beta} = (\slash\!\!\! p)_{\beta\alpha} +
\delta_{\beta\alpha}\,(1/2) \bar{D}D, \nonumber\\
&&D_{\alpha}\bar{D}_{\beta} = (\slash\!\!\! p)_{\alpha\beta} -
\delta_{\alpha\beta}\, (1/2) \bar{D}D.
\label{DD}
\end{eqnarray}
A direct calculation then shows that $D_{\alpha}$
anticommute with $\bar{D}D$, $\{D_{\alpha},\bar{D}D\} =0$.
This is combined with an obvious relation $[D_{\alpha},
\{\bar{D}_{\alpha}, D_{\beta}\} ] = 0$ to yield 
\begin{equation}
\bar{D}_{\alpha}D_{\beta}D_{\alpha}=0. 
\end{equation}
Some further formulas useful in operator calculus are
\begin{eqnarray}
&&(\bar{D}D)^{2} =  4\, p^{2}, \ \bar{D} \gamma ^{\mu}D = 2\,
p^{\mu},\nonumber\\
&&2(\slash\!\!\! p D)_{\alpha} 
= D_{\alpha}\bar{D}D = -\bar{D}DD_{\alpha}.
\end{eqnarray}

\section{Heat-kernel in superspace}

In this appendix we outline the calculation of the heat-kernel 
$\Gamma = e^{\tau {\cal D}^{2}}$ with 
${\cal D}=- {1\over{2}} \bar{D}D + W''(\Phi_{\rm c})$, 
relevant to Eq.~(\ref{Aalpha}). 
Let us substitute ${\cal D}^{2}=p^{2}- {1\over{2}}\, \{\bar{D}D, W''\}
+(W'')^{2}$ into $\Gamma$, and expand it in powers of $W''$. 
The first-order correction reads
\begin{equation}
\Gamma^{(1)} = -{1\over{2}} \int_{0}^{\tau} ds\, e^{(\tau-s) p^{2}}
\{\bar{D}D, W''\}\, e^{s p^{2}},
\end{equation}
where $W''=W''[\Phi_{\rm c}]$.
On taking the $\theta$-diagonal element,
one finds $(\theta|-{1\over{2}} \{\bar{D}D, W''\}|\theta)= 2W''$.
Hence 
\begin{eqnarray}
\langle z|\Gamma^{(1)}|z\rangle
&=& 2\langle x|e^{\tau p^{2}}\int_{0}^{\tau} ds\, 
e^{-s p^{2}} W'' e^{s p^{2}}|x\rangle \nonumber\\
&=& 2\tau\langle x|e^{\tau p^{2}} W''|x\rangle +\cdots \nonumber\\
&=& {i\over{2\pi}}\,W''(\Phi_{\rm c})\ \ \ \ (\tau \rightarrow 0_{+}),
\label{heatkernelGamma}
\end{eqnarray}
where we have used 
\begin{equation}
\langle x|e^{\tau p^{2}}|x\rangle 
= \int {d^{2}p\over{(2\pi)^{2}}} \, e^{\tau p^{2}}
={i\over{4\pi\tau}}.
\end{equation}
It is readily seen on dimensional grounds that the desired 
$\langle z|\Gamma|z\rangle$ is given by this first-order
result~(\ref{heatkernelGamma}) alone in the $\tau \rightarrow 0_{+}$
limit.

Finally we quote some other examples of anomalous products (in operator
form):
\begin{eqnarray}
(\bar{D}_{\alpha}\Phi) D_{\beta}{\delta S\over{\delta \Phi}}
&=& {1\over{4\pi}}\,(W'')^{2}\, \delta_{\alpha\beta}, \\
(\bar{D}D\Phi) {\delta S\over{\delta \Phi}}
&=&  -{1\over{4\pi}}\,\Big[ \bar{D}DW'' + 2(W'')^{2}\Big], \\
G(\Phi)\, {\delta S\over{\delta \Phi}}\
&=&  -{1\over{2\pi}}\, G'(\Phi)W''(\Phi) ,\label{GphiS}
\end{eqnarray}
where $G(\Phi)$ is a polynomial of $\Phi$.
These imply that 
$\bar{\psi}\,(\delta S/\delta\bar{\psi}) \not=0$, 
$F(\delta S/\delta F) = \phi\, (\delta S/\delta \phi) \not= 0$, etc.

\section{Propagator}

In this appendix we outline the calculation of the propagator
$\langle  \chi(z) \chi (z) \rangle 
= i\langle z| ( -{1\over{2}}\bar{D}D + M)^{-1} |z \rangle$ 
with $M = W''(\Phi_{c})$ in Eq.~(\ref{chipropagator}). 
Let us rewrite the propagator in the form
\begin{equation}
\langle z|{i\over{ p^{2} -(\bar{D}M)D -\kappa^{2}}} 
( -{1\over{2}}\bar{D}D - M) |z\rangle
\end{equation}
with  $\kappa^{2}= {1\over{2}} (\bar{D}D M) + M^{2}$.
We then expand it in powers of $(\bar{D}M)D$, bring the derivatives
$D_{\alpha}$ to the very right-hand side, and note formulas such as
$(\theta|-{1\over{2}}\bar{D}D |\theta)=1$. This leads to
expressions in ordinary space-time ($x^{\mu}$), classified in
powers of $D_{\alpha}$ acting on $M$. 
Retaining terms to $O(D^{2})$ yields
\begin{equation}
\langle x|{i\over{p^{2}\! -\! \kappa^{2}}}|x \rangle 
+\langle x| {i\over{(p^{2}\! -\! \kappa^{2})^{3}}}|x\rangle 
M(\bar{D}M)(DM),
\end{equation}
which eventually leads to Eq.~(\ref{chipropagator}).


\begin{thebibliography}{99}
\bibitem{fn}
As seen from Eq.~(\ref{GphiS}) in Appendix B, a product
of the form $FG(\phi)$ with a polynomial $G(\phi)$ turns into 
$-W'G(\phi) - (1/2\pi) W'' G'(\phi)$; this shows clearly 
how the auxiliary field $F$ changes its role  in each field product.
\end{thebibliography}
\end{document}